\documentclass[
reprint,superscriptaddress,
amsmath,amssymb,
aps,
pra,
]{revtex4-1}

\usepackage{graphicx}
\usepackage[caption=false]{subfig}
\usepackage{dcolumn}
\usepackage{bm}
\usepackage{color}
\usepackage{siunitx}

\newcommand{\bra}[1]{\langle #1|}
\newcommand{\ket}[1]{|#1\rangle}

\newcommand{\cZero}{\ket{\mathcal{C}_\alpha^{(0\mathrm{mod}4)}}}
\newcommand{\bcZero}{\bra{\mathcal{C}_\alpha^{(0\mathrm{mod}4)}}}

\begin{document}



\title{Generating higher order quantum dissipation from lower order parametric processes}

\author{S.O. Mundhada}
\author{A. Grimm}
\author{S. Touzard}
\author{U. Vool}
\author{S. Shankar}
\author{M.H. Devoret}
\affiliation{Department of Applied Physics, Yale University, New Haven, Connecticut 06520, USA}
\author{M. Mirrahimi}
\affiliation{QUANTIC team, INRIA de Paris, 2 Rue Simone Iff, 75012 Paris, France}
\affiliation{Yale Quantum Institute, Yale University, New Haven, Connecticut 06520, USA}

\date{\today}
\begin{abstract}
Stabilization of quantum manifolds is at the heart of error-protected quantum information storage and manipulation. Nonlinear driven-dissipative processes achieve such stabilization in a hardware efficient manner. Josephson circuits with parametric pump drives implement these nonlinear interactions. In this article, we propose a scheme to engineer a four-photon drive and dissipation on a harmonic oscillator by cascading experimentally demonstrated two-photon processes. This would stabilize a four-dimensional degenerate manifold in a superconducting resonator. We analyze the performance of the scheme using numerical simulations of a realizable system with experimentally achievable parameters.
\end{abstract}
\maketitle

\section{Introduction}

\begin{figure*}[t]
\centering
\subfloat{{\label{fig:basic_principle_a-}}
	\includegraphics[width=0.4\textwidth]{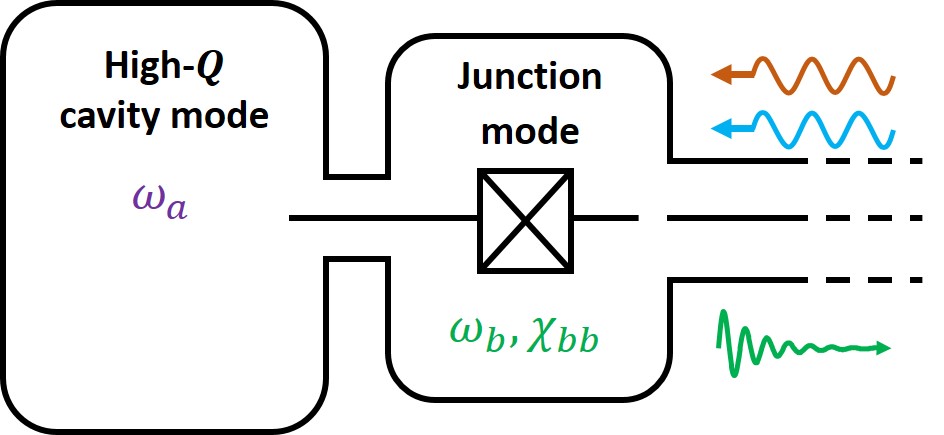}}
	\put(-210,95){a.}
\subfloat{{\label{fig:basic_principle_a}}
	\raisebox{0.1cm}{
	\includegraphics[width=0.4\textwidth]{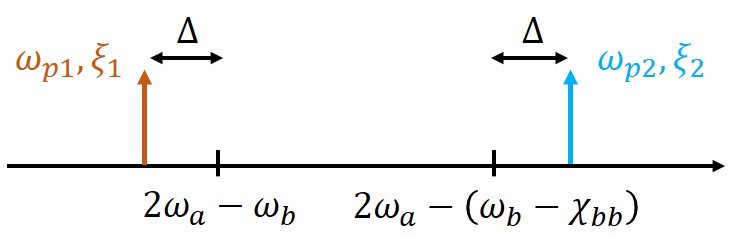}}}
	\put(-210,95){b.}\\
\subfloat{{\label{fig:basic_principle_b}}
	\includegraphics[width=0.4\textwidth]{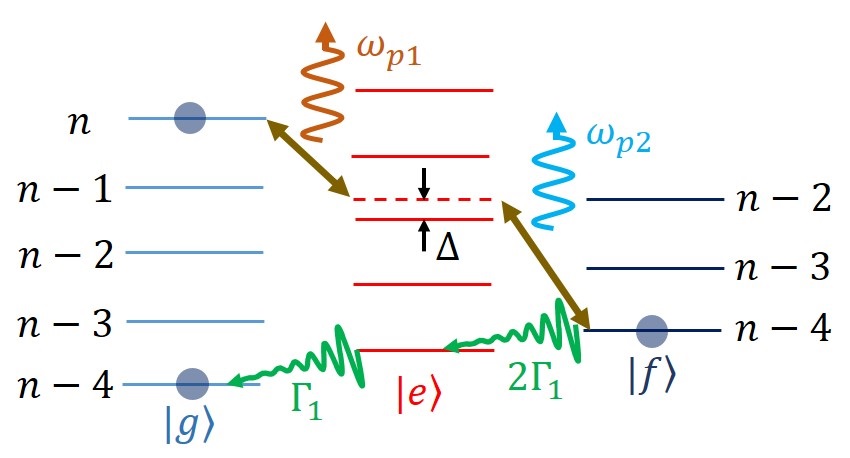}}
	\put(-210,95){c.}
\subfloat{{\label{fig:basic_principle_c}}
	\raisebox{0.6cm}{
	\includegraphics[width=0.4\textwidth]{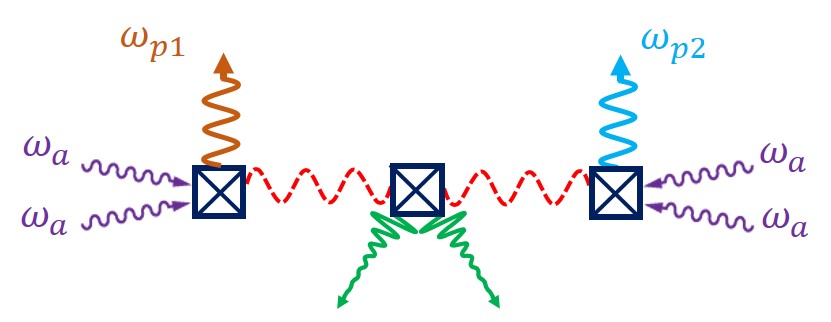}}}
	\put(-210,95){d.}
\caption{Basic principle behind cascading two-photon exchange processes. (a) A high-Q cavity at frequency $\omega_a$ coupled to a Josephson junction mode at frequency $\omega_b$ and anharmonicity $\chi_{bb}$. The system can be driven through a transmission line coupled to the junction mode as shown. (b) Driving the system at frequency $2\omega_a-\omega_b$ ($2\omega_a-\omega_b+\chi_{bb}$) would lead to an exchange of two cavity photons with the $g-e$ ($e-f$) excitations of the junction mode. Instead, by detuning the drives by $\Delta$ (respectively $-\Delta$) only a cascade of two such exchanges is possible, leading to a four-photon exchange. (c) Explanation of the cascading process using the energy level description of the junction-cavity system. Here, the Fock-states of the cavity are denoted by numbers and the lowest three eigenstates of the junction mode are denoted by letters $g$, $e$ and $f$. The first pump (brown) connects the state $\ket{g,n}$ with a virtual state detuned from state $\ket{e,n-2}$ by $\Delta$ (red dashed line). The second pump (cyan) connects this virtual state with the state $\ket{n-4,f}$. Thus a pair of two-photon exchanges are combined to create a four-photon transition from $\ket{g,n}$ to $\ket{f,n-4}$. (d) Diagramatic representation of the four photon exchange process. Two pairs of cavity photons (purple) each combine with a pump photon to produce a virtual junction excitation (dashed red). The resulting two virtual excitations in turn combine to create two real junction photons (green). \label{fig:basic_principle}}
\end{figure*}

In order to achieve a robust encoding and processing of quantum information, it is important to stabilize not only individual quantum states, but the entire manifold spanned by their superpositions. This requires synthesizing  artificial interactions with desirable properties which are impossible to find in natural systems. Particularly, in the case of quantum superconducting circuits, Josephson junctions together with parametric pumping methods provide powerful hardware elements for the design of such Hamiltonians. Here we extend the design toolkit, by using ideas borrowed from Raman processes to achieve Hamiltonians of high-order nonlinearity. More precisely, we introduce a novel nonlinear driven-dissipative process stabilizing a four-dimensional degenerate manifold. 

In general, a quantum system interacting with its environment will decohere through the entanglement between the environmental and the system degrees of freedom. However, in certain cases, a driven system with a properly tailored interaction with an environment can remain in a pure excited state or even a manifold of excited states. The simplest example is a driven harmonic oscillator with an ordinary dissipation, i.e. a frictional force proportional to velocity. In the underdamped quantum regime, this friction corresponds to the harmonic oscillator undergoing a single-photon loss process.  Such a driven-dissipative process, in the rotating frame of the harmonic oscillator, can be modeled by the master equation 
$$
\frac{\mathrm{d}}{\mathrm{d}t}\rho=-i[\epsilon_d \hat{a}^\dag+\epsilon_d^* \hat{a},\rho]+\kappa\mathcal{D}\left[\hat{a}\right]\rho,
$$
where $\rho$ is the density operator, $\hat a$ is the harmonic oscillator annihilation operator, $\epsilon_d$ represents the resonant complex amplitude of the resonant drive, $\kappa$ is the dissipation rate of the harmonic oscillator and 
$$
\mathcal{D}\left[\hat{L}\right]\rho=\hat{L}\rho\hat{L}^\dagger - \frac{1}{2}\hat{L}^\dagger \hat{L} \rho - \frac{1}{2} \rho \hat{L}^\dagger \hat{L} 
$$
is the Lindblad super-operator. The system admits a pure steady state, which is a coherent state denoted by $\ket{\alpha}$ where $\alpha=-2i\epsilon_d/\kappa$. Note also that the right-hand side of the above master equation can be simply written as $\kappa\mathcal{D}\left[\hat{a}-\alpha\right]\rho$. The fact that $\ket{\alpha}$ is the  steady state of the process follows from $(\hat{a}-\alpha)\ket{\alpha}=0$. This idea can be generalized to a non-linear dissipation  of the form $\kappa\mathcal{D}[a^n-\alpha^n]\rho$ which admits as steady states the $n$ coherent states $\{\ket{\alpha e^{2im\pi/n}}\}_{m=0}^{n-1}$. Indeed, all these coherent states and their superpositions are in the kernel of the dissipation operator $(a^n-\alpha^n)$. Therefore, this process stabilizes the whole $n$-dimensional manifold spanned by the above coherent states. 

The case with $n=2$ has been proposed in~\citep{Wolinsky1988,Gilles1994,HachIII1994,Mirrahimi2014} and experimentally realized in~\citep{Leghtas2015}. The idea consists of mediating a coupling between a high-Q cavity mode (resonance frequency $\omega_a$) and a low-Q resonator (resonance frequency $\omega_b$) through a Josephson junction. Applying a strong microwave drive at frequency $\omega_{\mathrm{pump}}=2\omega_a-\omega_b$ and a weaker drive at frequency $\omega_b$, we achieve an effective interaction Hamiltonian of the form 
$$
\frac{H_{\mathrm{2ph}}}{\hbar}=(g_{\mathrm{2ph}}^*\hat a^{\dag~2}\hat b+g_{\mathrm{2ph}}\hat a^{2}\hat b^\dag)-(\epsilon_d^* \hat b+\epsilon_d \hat b^\dag).
$$ 
Combining this interaction with a strong dissipation $\Gamma\mathcal{D}[\hat{b}]$ at the rate $\Gamma\gg|g_{\mathrm{2ph}}|$ translates to an effective dissipation of the form $\kappa_{\mathrm{2ph}}\mathcal{D}[\hat a^2-\alpha^2]\rho$, where $\kappa_{\mathrm{2ph}}=4|g_{\mathrm{2ph}}|^2/\Gamma$ and $\alpha=\sqrt{\epsilon_d/g_{\mathrm{2ph}}}$. Here we go beyond this by exploring a scheme which enables non-linear dissipations of higher-order. More precisely, we propose a method to achieve a four-photon interaction Hamiltonian without significantly increasing the required hardware complexity. The idea consists of using a Raman-type process~\citep{Gardiner2015}, exploiting virtual transitions, to cascade two $H_{\mathrm{2ph}}$ interactions.

An important  application of such a manifold stabilization is error-protected quantum information encoding and processing. As suggested in~\citep{Mirrahimi2014}, a four-photon driven-dissipative process enables a protected encoding of quantum information in two steady states of the same photon-number parity. Continuous monitoring of  the photon-number parity observable then enables a  protection against the dominant decay channel of the harmonic oscillator corresponding to the natural single-photon loss~\cite{Ofek2016}. 

Section~\ref{sec:cascading} describes the scheme to achieve a four-photon exchange Hamiltonian. In Section~\ref{sec:including_dissipation}, we study the dynamics in presence of dissipation of the low-Q mode along with possible improvements. In the Appendices~\ref{appendix:rwa_with_dissipation} and~\ref{appendix:optimizing_gamma} we discuss the derivation of the effective master equation and the accuracy of approximations used in the analytical calculations.

\section{Cascading nonlinear processes}
\label{sec:cascading}

Similar to the ideas presented in the last section, in order to have four-photon dissipation, we need to build a process that exchanges four cavity-photons with an excitation in a dissipative mode.  Following the example of the two photon process, this could be realized by engineering an interaction Hamiltonian of the form $H_{\mathrm{int}}/\hbar=g_{4\mathrm{ph}} \hat{a}^4 \hat{b}^\dagger + g_{4\mathrm{ph}}^* \hat{a}^{4 \dagger}\hat{b} $ which is exchanging four photons of the cavity mode with a single excitation of the low-Q mode. We need the strength of the interaction $|g_{4\mathrm{ph}}|$ to significantly exceed the decay rate of the storage cavity mode $\hat{a}$.

The Hamiltonian of a Josephson junction provides us with a six-wave mixing process which combined with an off-resonant pump at frequency $\omega_p=4\omega_a-\omega_b$ could in principle produce such an interaction. However, this six-wave mixing process comes along with other nonlinear terms in the Hamiltonian which could be of the same or higher magnitude. In particular, as it has been explained in~\citep{Mirrahimi2014}, with the currently achievable experimental parameters, the cavity self-Kerr effect would be at least an order of magnitude larger than $|g_{4\mathrm{ph}}|$.

Reference~\citep{Mirrahimi2014} proposes a more elaborate Josephson circuit to realize a purer interaction Hamiltonian. This, however, comes at the expense of significant hardware development and might encounter other unknown experimental limitations. Here we propose an alternative approach, which is based on cascading two-photon exchange processes. This leads to significant hardware simplifications and could in principle be realized with current experimental setups~\citep{Leghtas2015}. In the next subsection, we give a schematic representation of the proposed protocol, which uses higher energy levels of the junction mode and a cascading based on Raman transition~\citep{Gardiner2015}.  In Subsection~\ref{sec:rwa_without_dissipation} we sketch a mathematical analysis based on the second order rotating wave approximation (RWA).  This is supplemented by numerical simulations comparing the exact and the approximate Hamiltonians.

\subsection{Four-photon exchange scheme}
\label{sec:basic_principle}
In order to combine a pair of two-photon exchange processes, we take advantage of the junction mode being a multilevel anharmonic system. The basic principle of our scheme is illustrated in Fig.~\ref{fig:basic_principle}. More precisely, we exchange four cavity photons with two excitations of the junction mode. This could be done in a sequential manner by exchanging, twice, two cavity photons with an excitation of the junction mode, once from $g$ to $e$ and then from $e$ to $f$. However, as will be seen later, populating the $e$ level of the junction mode leads to undesired decoherence channels for the cavity mode. Therefore we perform this cascading using a virtual transition through the $e$ level by detuning the two-photon exchange pumps. This is similar to a Raman transition in a three level system. 

Consequently, we apply two pumps at frequencies, $\omega_{p1}=2\omega_a-\omega_b-\Delta$ and $\omega_{p2}=2\omega_a-(\omega_b-\chi_{bb})+\Delta$ as shown in Fig.~\ref{fig:basic_principle_a}. Note that here we are considering the $\hat{b}$ mode to be a junction mode with frequency $\omega_b$ and anharmonicity $\chi_{bb}$. For this protocol to work we require $\chi_{bb}\gg\Delta$, as we will justify in the next section. Starting in the state $\ket{g,n}$ these pumps make a transition to the state $\ket{f,n-4}$, passing virtually through the state $\ket{e,n-2}$ (see Fig.~\ref{fig:basic_principle_b}). As we will see in Section~\ref{sec:including_dissipation}, in order to achieve four-photon dissipation, we also require the junction mode to dissipate from the $f$ to the $g$ state. 

\subsection{Analytical derivation using second-order RWA}
\label{sec:rwa_without_dissipation}

\begin{figure}[t]
\centering
\subfloat{
	\includegraphics[width=0.45\textwidth]{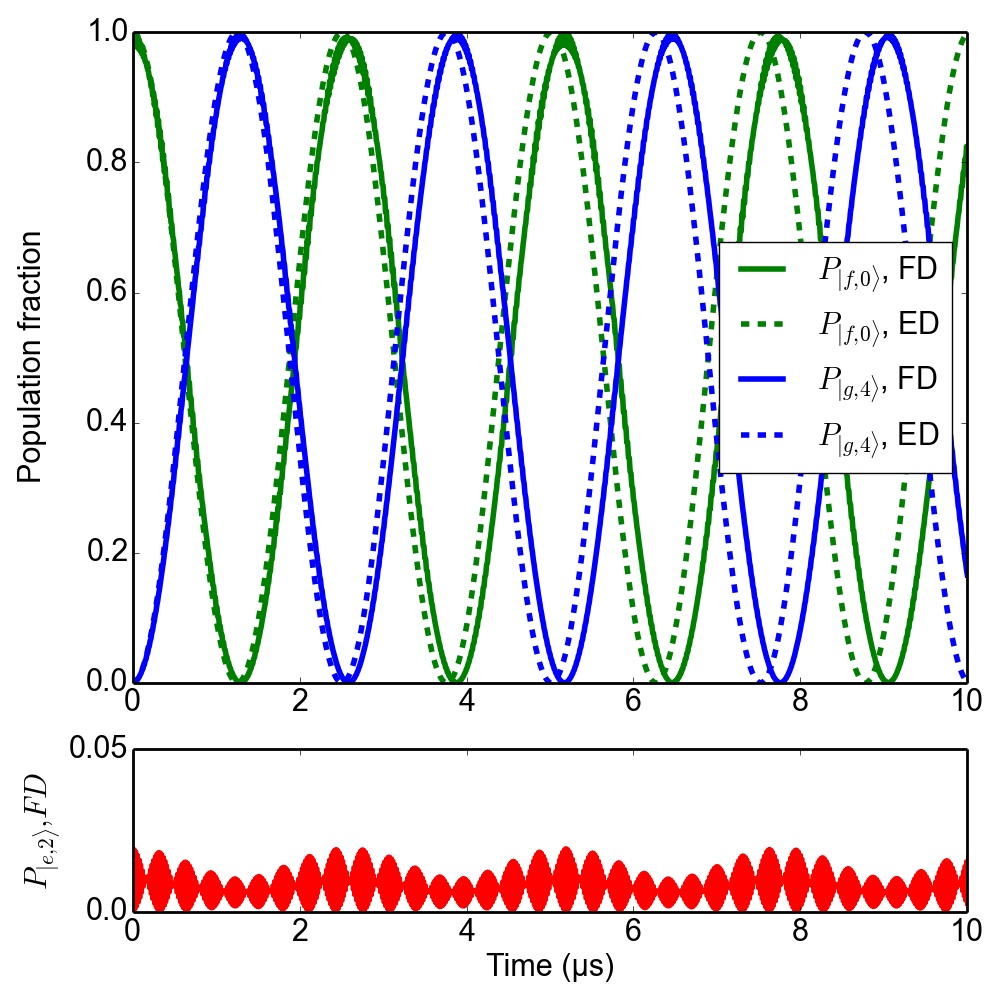}}
	\put(-230,220){a.}
	\put(-230,60){b.}
\caption{Numerical simulation of four-photon exchange process. (a) We compare the effective dynamics (ED) given by the Hamiltonian \eqref{eq:H_effective} with the full dynamics (FD) corresponding to \eqref{eq:H_sys}. For the four-photon exchange we do not need the $g\leftrightarrow f$ Rabi drive (see text). Consequently we set $g_3=0$ in these equations. We start in the state $\ket{f,0}$ and monitor the population of $\ket{g,4}$ (blue) and $\ket{f,0}$ (green). (b) Population leakage to the state $\ket{e,2}$. The system parameters are $\chi_{aa}/(2\pi)=\SI{312}{\hertz}$, $\chi_{bb}/(2\pi)=\SI{200}{\mega\hertz}$, $\chi_{ab}/(2\pi)=\SI{0.5}{\mega\hertz}$. We choose $\delta=\SI{153}{\kilo\hertz}$, $\Delta=\SI{50}{\mega\hertz}$, $g_1/(2\pi)=\SI{899}{\kilo\hertz}$, $g_2/(2\pi)=\SI{2}{\mega\hertz}$ in order to satisfy \eqref{eq:zeta_gg_constraint}. \label{fig:full_effective_comparison_H}}
\end{figure}

Here, starting from the full Hamiltonian of the junction-cavity system, we provide a mathematical analysis of the proposed scheme. We also include an additional drive in our calculations, which will address a two-photon transition between the $g$ and $f$ levels of the junction mode. The importance of this drive will be clear in Section~\ref{sec:including_dissipation} when we talk about a four-photon driven-dissipative process. The starting Hamiltonian is given by~\citep{Nigg2012}
\begin{align}
\frac{H(t)}{\hbar} =& \omega_a \hat{a}^\dagger \hat{a} + \omega_b \hat{b}^\dagger \hat{b} - \frac{E_J}{\hbar} \left[\cos{\left(\hat{\varphi}\right)}+\frac{\hat{\varphi}^2}{2!}\right] \nonumber\\
&+ \sum_{k=0}^3 \epsilon_{pk}(t) \left(\hat{b}+\hat{b}^\dagger\right), \label{eq:basic-hamiltonian}\nonumber
\end{align}
where $E_J$ is the Josephson energy and $\hat{\varphi}=\phi_a (\hat{a}+\hat{a}^\dagger)+\phi_b (\hat{b}+\hat{b}^\dagger)$. Here $\phi_{a(b)}=\phi_{\mathrm{ZPF},a(b)}/\phi_0$ with $\phi_{\mathrm{ZPF},a(b)}$ corresponds to the zero point fluctuations of the two modes as seen by the junction  and $\phi_0=\hbar/2e$ is the reduced superconducting flux quantum. The drive fields $\epsilon_{pk}(t)= 2\epsilon_{pk} \cos\left(\omega_{pk} t + \theta_{k}\right)$ represent the off-resonant pump terms. The pump frequencies are selected to be
\begin{align}
\omega_{p1}&=2\tilde\omega_a-\tilde\omega_b-\Delta+\delta \nonumber \\
\omega_{p2}&=2\tilde\omega_a-(\tilde\omega_b-\chi_{bb})+\Delta+\delta\nonumber\\
\omega_{p3}&= \tilde\omega_b - \frac{\chi_{bb}}{2} -\frac{\delta}{2}
\end{align}
where $\tilde{\omega}_{a}$ and $\tilde{\omega}_{b}$ are Lamb and Stark shifted cavity and junction mode frequencies. The additional detuning $\delta\ll \Delta$ will be selected to compensate for higher order frequency shifts. 

Following the supplementary material of~\citep{Leghtas2015}, we go into a displaced frame absorbing the pump terms in the cosine. This leads to the Hamiltonian
\begin{align}
\frac{H'(t)}{\hbar} =&  \omega_a \hat{a}^\dagger \hat{a} + \omega_b \hat{b}^\dagger \hat{b} - E_J \left[\cos{\left(\hat{\Phi}(t)\right)}+\frac{\hat{\Phi}^2(t)}{2!}\right], \nonumber
\end{align}
where
\begin{align}
\hat{\Phi}(t)= \phi_a \hat{a}+\phi_b \hat{b} +\phi_b \sum_{k=1}^3\xi_{k}\exp(-i\omega_{pk}t)+ \mathrm{h.c.}.\nonumber
\end{align}
Here $\mathrm{h.c.}$ stands for Hermitian conjugate and $\xi_{k}$ are complex coefficients related to phases and amplitudes of the pumps.

Developing the cosine up to fourth-order terms and keeping only the diagonal and the two-photon exchange terms, we get a Hamiltonian of the form
\begin{align}
\frac{H_{\mathrm{sys}}(t)}{\hbar}=&\tilde\omega_a \hat{a}^\dagger \hat{a} + \tilde\omega_b \hat{b}^\dagger \hat{b}  \nonumber\\
& - \frac{\chi_{aa}}{2} \hat{a}^{\dagger 2} \hat{a}^2 - \frac{\chi_{bb}}{2} \hat{b}^{\dagger 2} \hat{b}^2 - \chi_{ab} \hat{a}^\dagger \hat{a} \hat{b}^\dagger \hat{b} \nonumber\\
&+\sum_{k=1,2} \left(g_k \exp\left(-i\omega_{pk}t)\right) \hat{a}^{\dagger 2} \hat{b} + \mathrm{h.c.}\right)\nonumber\\
&-\left(g_3 \exp\left(2i\omega_{p3}t)\right) \hat{b}^2 + \mathrm{h.c.}\right).\label{eq:H_sys}
\end{align}
Here we have ignored all the other terms assuming a sufficiently large frequency difference,  $|\tilde{\omega}_a-\tilde{\omega}_b|$, between the two modes. Indeed, in the rotating frame of $\tilde\omega_a \hat{a}^\dagger \hat{a} + \tilde\omega_b \hat{b}^\dagger \hat{b}$ these terms will be oscillating at significantly higher frequencies. In the above Hamiltonian,  $\chi_{aa}$, $\chi_{bb}$ and $\chi_{ab}$ are respectively the self-Kerr and cross-Kerr couplings between the junction mode and the cavity mode. Furthermore, $\tilde\omega_a$ and $\tilde\omega_b$ are given by 
\begin{align}
\tilde\omega_a &= \omega_a-\chi_{aa}-\frac{\chi_{ab}}{2}-\chi_{ab}\sum_{k=1}^3|\xi_k|^2\nonumber\\
\tilde\omega_b &= \omega_b-\chi_{bb}-\frac{\chi_{ab}}{2}-2\chi_{bb}\sum_{k=1}^3|\xi_k|^2. \nonumber
\end{align}
Finally, the two photon exchange strengths $g_k$ are given by
\begin{align}
g_{1/2} &=  -\frac{\chi_{ab}}{2} \xi_{1/2} \hspace{0.5cm}\mathrm{and}\hspace{0.5cm} g_3 = \frac{\chi_{bb}}{2}\xi_3^{*2}. \nonumber
\end{align}

Going into rotating frame with respect to $H_0/\hbar = \tilde\omega_a \hat{a}^\dagger \hat{a} + (\tilde\omega_b-\delta) \hat{b}^\dagger \hat{b}  -\frac{\chi_{bb}}{2}  \hat{b}^{\dagger 2} \hat{b}^2$, the Hamiltonian becomes
 \begin{align}
\frac{ H_{\mathrm{I}}(t)}{\hbar}=& \delta \hat{b}^\dagger \hat{b} - \frac{\chi_{aa}}{2} \hat{a}^{\dagger 2} \hat{a}^2- \chi_{ab} \hat{a}^\dagger \hat{a} \hat{b}^\dagger \hat{b} \nonumber\\
&+ g_1 \exp{\left[i\left(\chi_{bb}\hat{b}^\dagger \hat{b}+\Delta\right)t\right]}  \hat{a}^{\dagger 2} \hat{b} + \mathrm{h.c.}\nonumber\\
&+ g_2 \exp{\left[i\left(\chi_{bb}\left(\hat{b}^\dagger \hat{b}-1\right)-\Delta\right)t\right]} \hat{a}^{\dagger 2} \hat{b} + \mathrm{h.c.}\nonumber\\
&- g_3 \exp{\left[2i\chi_{bb}\hat{b}^\dagger\hat{b}\right]}\hat{b}^2+ \mathrm{h.c.}\nonumber
\end{align}
As outlined in~\citep{Mirrahimi2015}, we perform second order RWA to get 
\begin{align}
H_{\mathrm{eff}} =& \overline{H_{\mathrm{I}}(t)}- i \overline{\left(H_{\mathrm{I}}(t)-\overline{H_{\mathrm{I}}(t)}\right)\int \mathrm{d}t\left(H_{\mathrm{I}}(t)-\overline{H_{\mathrm{I}}(t)}\right)} \nonumber
\end{align}
where $\overline{A(t)}=\lim_{T\to\infty}\frac{1}{T}\int_0^T A(t)\mathrm{d}t$. Using the expression for $H_{\mathrm{I}}(t)$, we get 
\begin{multline}
\frac{H_{\mathrm{eff}}}{\hbar}= \left(g_{4\mathrm{ph}} \hat{a}^{\dagger 4} -\epsilon_{4\mathrm{ph}}\right) \hat\sigma_{fg} + \mathrm{h.c.}\\
 + \left(\zeta_{gaa} \hat\sigma_{gg} + \zeta_{eaa} \hat\sigma_{ee} + \zeta_{faa}\hat\sigma_{ff}-\frac{\chi_{aa}}{2}\right)  \hat{a}^{\dagger 2} \hat{a}^2 \\
 + \left((\chi_{ea}-\chi_{ab}) \hat\sigma_{ee} + (\chi_{fa}-2\chi_{ab}) \hat\sigma_{ff} \right) \hat{a}^\dagger \hat{a} \\
 + \left(\delta+\frac{\chi_{ea}}{2}-\frac{3|g_3|^2}{\chi_{bb}}\right)\hat\sigma_{ee} +\left(2\delta+\frac{\chi_{fa}}{2}\right) \hat\sigma_{ff} \label{eq:H_effective}
\end{multline}
where we have only considered the first three energy levels $g$, $e$ and $f$ of the junction mode. The other energy levels of this mode are never populated in this scheme. The transition operators $\hat\sigma_{jk}$ are given by $\ket{k}\bra{j}$. The first row of \eqref{eq:H_effective} is the four-photon exchange term and the two-photon $g\leftrightarrow f$ drive on the junction mode with
\begin{align*}
g_{4\mathrm{ph}} = \sqrt{2} g_1 g_2\left(\frac{1}{\Delta}-\frac{1}{\chi_{bb}+\Delta}\right) \mbox{ and }\epsilon_{4\mathrm{ph}} = \sqrt{2}g_3.
\end{align*}
In addition to this, the pumping also modifies the cross-Kerr terms by
\begin{align*}
\chi_{ea} =& \frac{4|g_2|^2}{\chi_{bb}-\Delta} - \frac{4|g_1|^2}{\Delta},\nonumber\\
\chi_{fa} =& \frac{8|g_2|^2}{\Delta}- \frac{8|g_1|^2}{\chi_{bb}+\Delta}
\end{align*}
and produces higher order interactions
\begin{align*}
\zeta_{gaa} =& \left(\frac{|g_1|^2}{\Delta}-\frac{|g_2|^2}{\chi_{bb}+\Delta}\right),\nonumber\\
\zeta_{eaa} =& \left(-\frac{|g_1|^2(\chi_{bb}-\Delta)}{\Delta(\chi_{bb}+\Delta)}-\frac{|g_2|^2(2\chi_{bb}+\Delta}{\Delta(\chi_{bb}+\Delta)}\right),\nonumber\\
\zeta_{faa}   =& \left(\frac{|g_2|^2(2\chi_{bb}+\Delta)}{\Delta (\chi_{bb}-\Delta)}-\frac{|g_1|^2}{2\chi_{bb}+\Delta}\right) .
\end{align*}

In order to show the correctness of the effective dynamics, let us consider the oscillations between the states $\ket{f,0}$ and $\ket{g,4}$. Note that the population of the state $\ket{e,n-2}$ will remain small. The terms $\left(\zeta_{gaa}\hat{\sigma}_{gg}-\chi_{aa}/2\right)\hat{a}^{\dagger 2}\hat{a}^2$ and $(2\delta+\chi_{fa}/2)\hat{\sigma}_{ff}$ produce additional frequency shifts between $\ket{g,4}$ and $\ket{f,0}$, thus hindering the oscillations. We counter the effect of these terms by selecting parameters such that 
\begin{align}
\zeta_{gaa}=\frac{\chi_{aa}}{2} \mbox{ and } \delta=-\frac{\chi_{fa}}{4}. \label{eq:zeta_gg_constraint}
\end{align}
The dynamics given by Hamiltonian \eqref{eq:H_sys} (simulated in the rotating frame of $\tilde{\omega}_a\hat{a}^\dagger \hat{a}+\tilde{\omega}_b\hat{b}^\dagger \hat{b}$) is compared with the effective dynamics given by Hamiltonian \eqref{eq:H_effective} in Fig.~\ref{fig:full_effective_comparison_H}a. The system parameters are $\chi_{aa}/(2\pi)=\SI{312}{\hertz}$, $\chi_{bb}/(2\pi)=\SI{200}{\mega\hertz}$, $\chi_{ab}/(2\pi)=\SI{0.5}{\mega\hertz}$ satisfying $\chi_{ab}=2\sqrt{\chi_{aa}\chi_{bb}}$~\citep{Nigg2012}. The values $\Delta/(2\pi)=\SI{50}{\mega\hertz}$, $\delta=\SI{153}{\kilo\hertz}$, $g_1/(2\pi)=\SI{899}{\kilo\hertz}$ and $g_2/(2\pi)=\SI{2}{\mega\hertz}$ are selected to satisfy \eqref{eq:zeta_gg_constraint}. The third drive $g_3$ is set to zero in this simulation. Dynamics given by both, \eqref{eq:H_sys} and \eqref{eq:H_effective}, show the required oscillations. The slight mismatch between the oscillation frequencies is due to a higher order effect induced by the occupation of the state $\ket{e,2}$. Figure~\ref{fig:full_effective_comparison_H}b shows the population leakage to the $\ket{e,2}$ state. This leakage leads to an important limitation of the protocol (see Subsection~\ref{sec:effective_master_equation}). 

\section{Four-photon driven-dissipative process}
\label{sec:including_dissipation}

We have showed in the last section that we get a four-photon exchange Hamiltonian \eqref{eq:H_effective} by cascading two-photon exchange processes. In this section we combine this idea with the dissipation of the junction mode to achieve a four-photon driven-dissipative dynamics on the cavity mode. In Subsection~\ref{sec:effective_master_equation}, we present the effective master equation governing the dynamics of the cavity. In particular, we observe that, as an undesired effect of population leakage towards the $e$ state, we introduce a two-photon dissipation on the cavity mode. This problem is addressed in Subsection~\ref{sec:two_photon_dissipation_error} by engineering the noise spectral density seen by the junction mode. Additionally, we analyze the performance of the proposed schemes through numerical simulations of the full and effective master equations.

\subsection{Effective master equation}
\label{sec:effective_master_equation}

\begin{figure*}[t]
\centering
\subfloat{{\label{fig:fidelity_purity}}
	\includegraphics[width=0.75\textwidth]{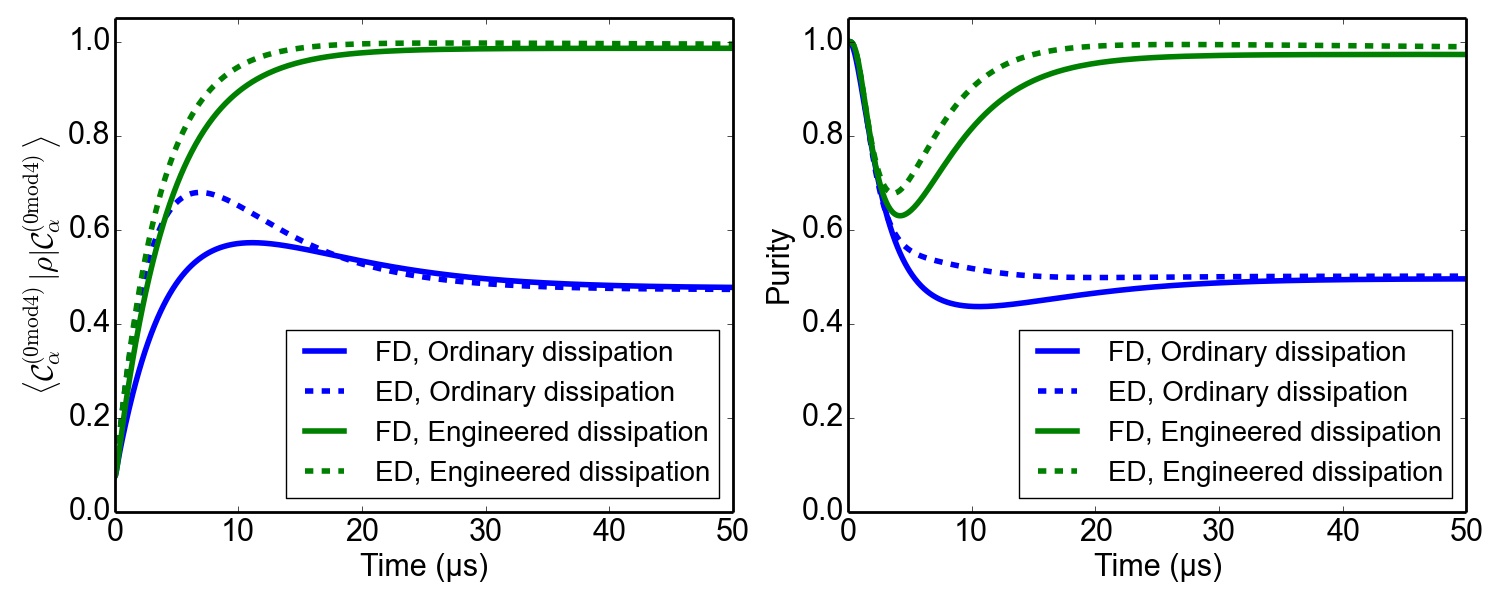}}
	\put(-365,150){a.}
	\put(-180,150){b.}
\subfloat{{\label{fig:wigners_vs_time}}
	\includegraphics[width=0.25\textwidth]{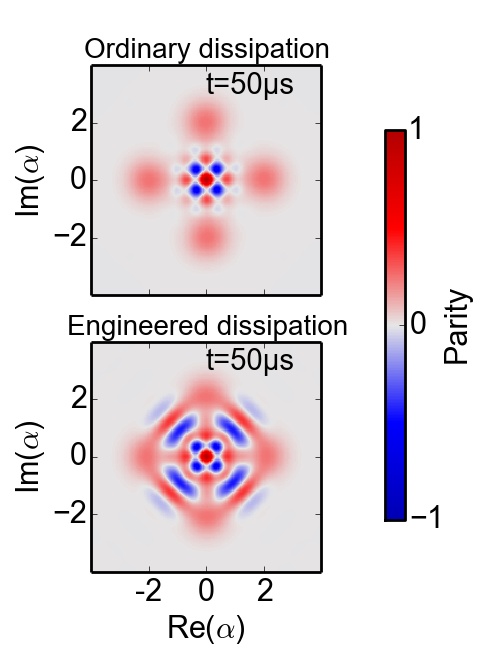}}
	\put(-102,142){c.}
	\put(-102,70){d.}
\caption{Simulation results for the full dynamics (FD, solid lines) given by the master equation \eqref{eq:full_eom_gamma12} and for the effective master equation \eqref{eq:effective_eom} (ED, dashed lines) with $\kappa_{4ph}$ and $\kappa_{2ph}$ given by \eqref{eq:improved_scheme_rates} and \eqref{eq:kappa_4ph_kappa_2ph} respectively. The blue curves correspond to a white noise spectrum (ordinary dissipation) described in Section~\ref{sec:effective_master_equation} and the green curves illustrate the result for an engineered $f$ to $g$ dissipation as in Section~\ref{sec:two_photon_dissipation_error}. Panel (a) presents the overlap of the density matrix with the state $\cZero$ and panel (b) plots the purity of the cavity state. Panel (c) and panel (d) show the Wigner functions at $t=50 \mu$s. The superior performance of the engineered dissipation is apparent from the results. The simulation parameters are $\chi_{bb}/(2\pi)=\SI{200}{\mega\hertz}$, $\chi_{aa}/(2\pi)=\SI{312}{\hertz}$,  $\chi_{ab}/(2\pi)=\SI{0.5}{\mega\hertz}$, $\Delta/(2\pi)=\SI{50}{\mega\hertz}$, $g_2/(2\pi)=\SI{2}{\mega\hertz}$ and $g_1/(2\pi)=\SI{899}{\kilo\hertz}$ such that $\zeta_{gaa}=\chi_{aa}/2$. For the ordinary dissipation $\Gamma_1/(2\pi)=\SI{2}{\mega\hertz}$, and for the engineered dissipation $\Gamma_1/(2\pi)=\SI{3}{\kilo\hertz}$ with the $f\rightarrow g$ direct dissipation given by $\Gamma_{fg}^{\mathrm{eng}}/(2\pi)=\SI{4}{\mega\hertz}$. \label{fig:simulation_results}}
\end{figure*}

We consider the junction mode to be coupled to a cold bath, leading to the master equation
\begin{align}
\frac{\mathrm{d}}{\mathrm{d}t}\rho = -\frac{i}{\hbar}\left[H_{\mathrm{sys}}(t), \rho\right]+\Gamma_1\mathcal{D}[\hat{b}]\rho. \label{eq:full_eom}
\end{align}
where the Hamiltonian $H_{\mathrm{sys}}$ is given by~\eqref{eq:H_sys}. Note that this master equation implicitly assumes a  white noise spectrum for the bath degrees of freedom. In Appendix~\ref{appendix:rwa_with_dissipation}, we will provide a more general analysis considering an arbitrary noise spectrum.  Indeed, in this appendix, we perform RWA under such general assumptions, arriving at a time-independent master equation for the junction-cavity system. Under the assumption of strong dissipation, we can also eliminate the junction degrees of freedom, resulting in an effective master equation for the cavity mode (see Appendix~\ref{appendix:optimizing_gamma}):
\begin{align}
\frac{\mathrm{d}}{\mathrm{d}t}\rho_{\mathrm{cav}}=& -i\left[\left(\zeta_{gaa}-\chi_{aa}\right)\hat{a}^{\dagger 2}\hat{a}^2,\rho_{\mathrm{cav}}\right] \nonumber\\
&+ \kappa_{4\mathrm{ph}} \mathcal{D}[\hat{a}^4-\alpha^4]\rho_{\mathrm{cav}} + \kappa_{2\mathrm{ph}} \mathcal{D}[\hat{a}^2]\rho_{\mathrm{cav}} \label{eq:effective_eom}
\end{align} 
with
\begin{align}
\kappa_{4\mathrm{ph}} =& \frac{2 |g_{4\mathrm{ph}}|^2}{\Gamma_1},\nonumber\\
\kappa_{2\mathrm{ph}} =& \left(\frac{|g_1|^2}{\Delta^2}+\frac{|g_2|^2}{(\Delta+\chi_{bb})^2}\right)\Gamma_1,\nonumber \\
\alpha =& \left(\frac{\epsilon_{4\mathrm{ph}}}{g_{4\mathrm{ph}}}\right)^{1/4}. \label{eq:kappa_4ph_kappa_2ph}
\end{align}
While we get the expected four-photon driven-dissipative term $\kappa_{4\mathrm{ph}}\mathcal{D}[\hat{a}^4-\alpha^4]$, we also inherit an undesired two-photon dissipation $\kappa_{2\mathrm{ph}}\mathcal{D}[\hat{a}^2]$. Such two photon dissipation corresponds to jumps between states with same photon number parities thus effectively introducing bit-flip errors in the logical code-space \citep{Mirrahimi2014}. In the next subsection, we will remedy this problem by engineering the dissipation of the junction mode. 

To establish the validity of \eqref{eq:effective_eom}, we numerically compare the dynamics of the two master equations~\eqref{eq:full_eom} and~\eqref{eq:effective_eom}. The blue curves in Fig.~\ref{fig:simulation_results}a and~\ref{fig:simulation_results}b correspond to these simulations. We initialize the system in its ground state and plot the overlap with the cat state $\cZero=\mathcal{N}\left(\ket{\alpha}+\ket{-\alpha}+\ket{i\alpha}+\ket{-i\alpha}\right)$ where $\mathcal{N}$ is a normalization factor. Note that as the cavity is initialized in the vacuum state, we expect the four-photon driven-dissipative process to steer the state towards $\cZero$~\citep{Mirrahimi2014}.  The chosen system parameters are the same as in the last section. The additional dissipation parameter $\Gamma_1/(2\pi) = \SI{2}{\mega\hertz}$. We also select $g_3/(2\pi)=\SI{460}{\kilo\hertz}$ to achieve a cat amplitude of $\alpha=2$. These parameters give $1/\kappa_{4\mathrm{ph}}\sim \SI{96}{\micro\second}$ and $1/\kappa_{2\mathrm{ph}}=\SI{205}{\micro\second}$. The maximum achieved overlap with the target state ($\cZero$) is merely above $50\%$. This is expected, since the two-photon dissipation rate is not much smaller than the four-photon dissipation rate. More precisely, the resulting steady state is a mixture of two even parity states represented by the Wigner function in Fig.~\ref{fig:simulation_results}c. Note that while this is a mixed state, the conservation of the photon number parity leads to negative values in the Wigner function. 

\subsection{Mitigation of two-photon dissipation error}
\label{sec:two_photon_dissipation_error}

As mentioned in the previous subsection, the inherited two-photon dissipation can be seen as a bit-flip error channel in the code space. Its rate has to be compared to the rate of other errors that are not corrected by the four-component cat code. Indeed, this code can only correct for a single photon loss  in the time interval $\delta t$ between two error syndrome (photon-number parity) measurements. The probability 
of two single-photon losses during $\delta t$ is given by $ p_{1}^{2\mathrm{ph}}=(|\alpha|^2\kappa_{1ph}\delta t)^2/2$. Whereas the probability for a direct two-photon loss due to the $\hat a^2$ dissipation  is $p_{2}^{2\mathrm{ph}}=|\alpha|^4\kappa_{2\mathrm{ph}}\delta t$. Hence, we require the induced error probability $p_2^{2\mathrm{ph}}$ to be of the same order or smaller than  $p_1^{2\mathrm{ph}}$. Therefore, we need to reduce $\kappa_{2\mathrm{ph}}$ to a value smaller than $\kappa_{1\mathrm{ph}}^2\delta t/2$. In this subsection,  we propose a simple modification of the above scheme, which, with currently achievable experimental parameters, should lead to $\kappa_{2\mathrm{ph}}/\kappa_{1\mathrm{ph}}$ to be less than $0.01$. 

One such approach is to use a dynamically engineered coupling of the junction mode to the bath. More precisely, we start with a high-Q junction mode corresponding to a small $\Gamma_1$ with respect to the Hamiltonian parameters. By dispersively coupling this mode to a low-Q resonator in the photon-number resolved regime~\citep{Schuster2007}, one can engineer a dynamical cooling protocol similar to DDROP~\citep{Geerlings2013}, or parametric sideband cooling~\citep{Pechal2014,Narla2016}. While these experiments correspond to a dynamical cooling from $e$ to $g$, one can easily modify them to achieve  a direct dissipation from $f$ to $g$. Here we model this engineered dissipation by adding a Lindblad term of the form $\Gamma_{fg}^{\mathrm{eng}}\mathcal{D}[\hat \sigma_{fg}]\rho$.  This leads to the new dissipation rate
\begin{equation}
\kappa_{4\mathrm{ph}} = \frac{4 |g_{4\mathrm{ph}}|^2}{\Gamma_{fg}^{\mathrm{eng}}+2\Gamma_1},\label{eq:improved_scheme_rates}
\end{equation}
while the two-photon dissipation rate $\kappa_{2\mathrm{ph}}$ remains unchanged, as given by~\eqref{eq:kappa_4ph_kappa_2ph}. 

The green curves in Fig.~\ref{fig:simulation_results}a and~\ref{fig:simulation_results}b illustrate the numerical simulations of this modified scheme. The solid curves correspond to the simulation of the master equation 
\begin{align}
\frac{\mathrm{d}}{\mathrm{d}t}\rho = -\frac{i}{\hbar}\left[H_{\mathrm{sys}}(t), \rho\right]+\Gamma_1\mathcal{D}[\hat{b}]\rho + \Gamma_{fg}^{\mathrm{eng}}\mathcal{D}[\hat{\sigma}_{fg}]\rho, \label{eq:full_eom_gamma12}
\end{align}
with $\Gamma_1/(2\pi)=\SI{3}{\kilo\hertz}$ and $\Gamma_{fg}^{\mathrm{eng}}/(2\pi)=\SI{4}{\mega\hertz}$. This value is a compromise between the strength of $\kappa_{4\mathrm{ph}}$ and the validity of the adiabatic elimination, as shown in Appendix~\ref{appendix:optimizing_gamma}. Similarly, the dashed green curves correspond to the simulation of \eqref{eq:effective_eom} with $\kappa_{4\mathrm{ph}}$ now given by \eqref{eq:improved_scheme_rates}. Indeed, with these parameters, we achieve $1/\kappa_{4\mathrm{ph}}\sim \SI{96}{\micro\second}$ and $1/\kappa_{2\mathrm{ph}}=\SI{136}{\milli\second}$. In Appendix~\ref{appendix:optimizing_gamma}, we will provide an alternative approach based on using band-pass Purcell filters~\citep{Reed2010}, shaping the noise spectrum seen by the junction mode.

\section{Conclusion}
\label{sec:conclusion}

We have presented a theoretical proposal for the implementation of a controlled four-photon driven-dissipative process on a harmonic oscillator. By stabilizing the manifold span$\{\ket{\pm\alpha},\ket{\pm i\alpha}\}$, this process provides a means to realize an error-corrected logical qubit~\citep{Mirrahimi2014}. Our proposal relies on cascading two-photon exchange processes, which have already been experimentally demonstrated~\citep{Leghtas2015}. 
While the required hardware complexity is similar to the existing system, the parameters need to be carefully chosen to avoid undesired interactions.  

The technique of cascading nonlinear processes through Raman-like virtual transitions can be used to engineer other highly nonlinear interactions. In particular, a Hamiltonian of the form $g_{\mathrm{12}}\hat{a}_1^{2\dagger}\hat{a}_2^2+g^*_{12}\hat{a}_1^2 \hat{a}_2^{2\dagger}$  could entangle two logical qubits encoded in two high-Q cavities $\hat{a}_1$ and $\hat{a}_2$~\citep{Mirrahimi2014}. Such an interaction can be generated by coupling the cavities through a Josephson junction mode $\hat{b}$ and applying two off-resonant pumps at frequencies  $\omega_{p1} = 2\tilde\omega_{a1}-\tilde\omega_b-\Delta$ and $\omega_{p2}=2\tilde\omega_{a2}-\tilde\omega_b-\Delta$. This entangling gate constitutes another important step towards fault-tolerant universal quantum computation with cat-qubits. 

\section*{Acknowledgement}

We acknowledge fruitful discussions with Ananda Roy, Zlatko Minev and Richard Brierly. This research was supported by INRIA’s DPEI under the TAQUILLA associated team and by ARO under Grant No. W911NF-14-1-0011.

\appendix

\section{RWA in presence of dissipation}
\label{appendix:rwa_with_dissipation}

\begin{figure}[t]
\centering
\subfloat{\includegraphics[width=0.4\textwidth]{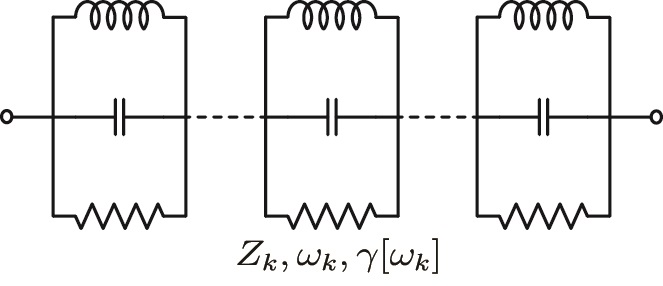}}
\caption{Modelling dissipation using LCR elements \cite{Solgun2014}. The system is assumed to be linearly coupled to infinitely many dissipative harmonic oscillators with finite frequency spacing. This can be seen as a discretization of a general noise spectrum using a basis of Lorentzian functions. \label{fig:foster}}
\end{figure}

We start by considering the Hamiltonian of a junction-cavity system where the junction mode is dissipative. This dissipation is typically modeled by a linear coupling to a continuum of infinitely many non-dissipative modes~\citep{Caldeira1983}. Here, instead, we model this dissipation by a linear coupling of the junction mode to infinitely many harmonic oscillators with finite frequency spacing and finite bandwidths. Indeed, for an under-coupled system (weak dissipation), we can use LCR elements~\citep{Nigg2012,Solgun2014} to represent the dissipation in terms of such dissipative oscillators (see Fig.~\ref{fig:foster}). Such a discretization could also be explained taking into account experimental considerations where the dissipation is mediated by various filters which could themselves be seen as lossy resonators. More precisely, we consider a Hamiltonian
\begin{align}
\frac{H_{\mathrm{tot}}}{\hbar} =& \frac{H_{\mathrm{sys}}}{\hbar}+ \sum_k \omega_k \hat{c}^{\dagger}[\omega_k]\hat{c}[\omega_k]\nonumber\\ 
&+ \sum_k \left(\Omega[\omega_k] \hat{b}\hat{c}^{\dagger}[\omega_k]+\Omega^*[\omega_k] \hat{b}^\dagger\hat{c}[\omega_k] \right), \nonumber
\end{align}
where $H_{\mathrm{sys}}$ is the system Hamiltonian given in \eqref{eq:H_sys} and the modes $\hat{c}[\omega_k]$ have decay rates $\gamma[\omega_k]$. We perform the second-order RWA on the associated master equation by going into the rotating frame of $\widetilde{H}_0 = \tilde\omega_a \hat{a}^\dagger \hat{a} + \tilde\omega_b \hat{b}^\dagger \hat{b} -\frac{\chi_{bb}}{2}  \hat{b}^{\dagger 2} \hat{b}^2 + \sum_k  \hbar\omega_k \hat{c}^\dagger[\omega_k] \hat{c}[\omega_k]$. The effective master equation becomes
\begin{align}
\frac{\mathrm{d}}{\mathrm{d}t}{\rho} =& -\frac{i}{\hbar}\left[H_{\mathrm{eff,bath}},\rho\right]+\sum_k (1+n_{\mathrm{th}}[\omega_k])\gamma[\omega_k]\mathcal{D}\left[\hat{c}[\omega_k]\right]\rho \nonumber\\
&+\sum_k n_{\mathrm{th}}[\omega_k]\gamma[\omega_k]\mathcal{D}\left[\hat{c}^\dagger[\omega_k]\right]\rho \nonumber
\end{align}
where $n_{\mathrm{th}}[\omega_k]$ implies the thermal population of the $k$th mode. The Hamiltonian $H_{\mathrm{eff,bath}}$ is given by
\begin{widetext}
\begin{align}
\frac{H_{\mathrm{eff,bath}}}{\hbar}\approx&\frac{H_{\mathrm{eff}}}{\hbar}+\sum_{n=0}^{\infty} \left(\Omega[\tilde\omega_b-(n-1)\chi_{bb}] \hat{c}^\dagger[\tilde\omega_b-\chi_{bb}]\sqrt{n}\hat\sigma_{n,n-1} +\mathrm{h.c.}\right)\nonumber\\
&+\left(g_1^* \Omega[\omega_{+\Delta,0}]\hat{c}^\dagger[\omega_{+\Delta,0}] \hat{a}^2  + \mathrm{h.c.}\right)\sum_{n=0}^{\infty}\frac{\chi_{bb}-\Delta}{(n\chi_{bb}+\Delta)((n-1)\chi_{bb}+\Delta)}\hat\sigma_{n,n} \nonumber\\
&+\left(g_2^* \Omega[\omega_{-\Delta,1}]\hat{c}^\dagger[\omega_{-\Delta,1}] \hat{a}^2  + \mathrm{h.c.}\right)\sum_{n=0}^{\infty}\frac{2\chi_{bb}+\Delta}{((n-2)\chi_{bb}-\Delta)((n-1)\chi_{bb}-\Delta)}\hat\sigma_{n,n} \nonumber\\
&+\sum_{n=0}^{\infty} \left(\frac{g_1\chi_{bb}\Omega[\omega_{+\Delta,2n}]}{(n\chi_{bb}+\Delta)((n+1)\chi_{bb}+\Delta)}\hat{c}^\dagger[\omega_{+\Delta,2n}]\hat{a}^{\dagger 2} \sqrt{(n+1)(n+2)}\hat{\sigma}_{n+2,n}+\mathrm{h.c.}\right)\nonumber\\
&+\sum_{n=0}^{\infty} \left(\frac{g_2\chi_{bb}\Omega[\omega_{-\Delta,2n+1}]}{(n\chi_{bb}-\Delta)((n-1)\chi_{bb}-\Delta)}\hat{c}^\dagger[\omega_{-\Delta,2n+1}] \hat{a}^{\dagger 2} \sqrt{(n+1)(n+2)}\hat{\sigma}_{n+2,n}+\mathrm{h.c.}\right).\label{eq:RWA_full_result}
\end{align}
\end{widetext}
Here $n$ indicates the number states of the junction mode (specifically $n=0,1,2$ correspond to $g,e,f$ levels in the main text). The frequencies $\omega_{\pm\Delta,n}$ are defined as $\tilde\omega_b\pm\Delta-n\chi_{bb}$. Along with the terms presented in \eqref{eq:RWA_full_result}, we also obtain terms of the form $\hat{\sigma}_{nn}\hat{c}^\dagger[\omega_k]\hat{c}[\omega_k]$ and $\hat{\sigma}_{n+2,n}\hat{c}^\dagger[\omega_k]\hat{c}^\dagger[\omega_m]+\mathrm{h.c.}$. The first type of terms corresponds to the dispersive coupling of the junction mode and the bath modes. For non-zero bath temperature, they contribute to the dephasing of the junction mode states. The latter terms become resonant when $\hbar(\omega_k+\omega_m)$ equals the energy difference between the states $n$ and $n+2$ of the junction mode and give rise to a direct two-photon dissipation between the two states. As stated in Section~\ref{sec:two_photon_dissipation_error}  such a direct dissipation from the $f$ to $g$ state actually enhances the performance of the protocol. However, without additional engineering, the magnitude of such interactions is negligible compared to the regular single-photon dissipation terms in \eqref{eq:RWA_full_result}.

\begin{figure*}[ht!]
\centering
\subfloat{
	\includegraphics[width=1.0\textwidth]{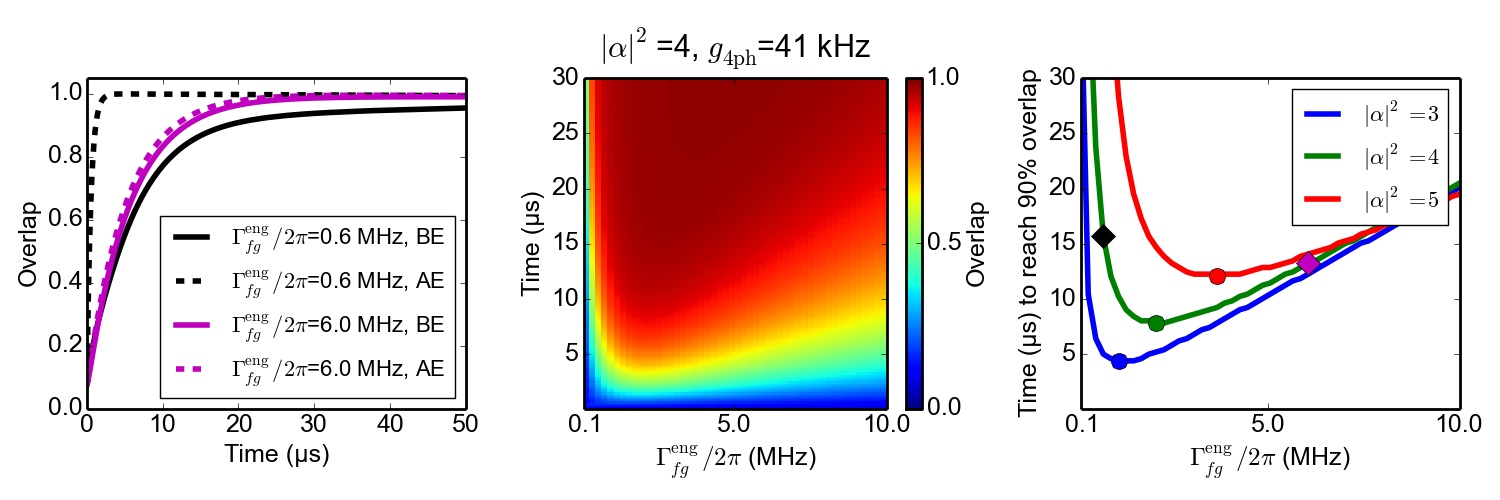}}
	\put(-500,155){a.}
	\put(-335,155){b.}
	\put(-155,155){c.}
\caption{Panel (a) compares the system dynamics before~\eqref{eq:before_elimination} and after~\eqref{eq:effective_eom} the adiabatic elimination, denoted respectively by (BE) and (AE).  We use the same parameters as in Fig.~\ref{fig:simulation_results} together with $\Gamma_{fg}^{\mathrm{eng}}/2\pi=\SI{0.6}{\mega\hertz}$ (black) and $\Gamma_{fg}^{\mathrm{eng}}/2\pi=\SI{6}{\mega\hertz}$ (magenta). We initialize the system in the ground state and plot the overlap  $\bcZero \rho \cZero$. Panel (b) shows the overlap as a function of time, obtained by simulating \eqref{eq:before_elimination} with $\Gamma_{fg}^{\mathrm{eng}}/2\pi$ ranging from $\SI{0.1}{\mega\hertz}$ to $\SI{10}{\mega\hertz}$. Panel (c) plots the time taken to achieve 90\% fidelity as a function of $\Gamma_{fg}^{\mathrm{eng}}$ for $|\alpha|^2=3,4$ and $5$. Black and magenta squares indicate the choices $\Gamma_{fg}^{\mathrm{eng}}/2\pi=\SI{0.6}{\mega\hertz}$ and $\Gamma_{fg}^{\mathrm{eng}}/2\pi=\SI{6}{\mega\hertz}$, corresponding to the simulations of (a). The dots correspond to the optimum working points for various cat amplitudes. }
\label{fig:adiabatic_validity} 
\end{figure*}

Next, by adiabatically eliminating the highly dissipative bath modes, we obtain the master equation
\begin{widetext}
\begin{align}
\frac{\mathrm{d}}{\mathrm{d}t}\rho =& -\frac{i}{\hbar}\left[H_{\mathrm{eff}},\rho\right]+\sum_{n=0}^{\infty} \left(n\Gamma_{\downarrow}[\tilde\omega_b-(n-1)\chi_{bb}]\mathcal{D}[\hat\sigma_{n,n-1}]+ n\Gamma_{\uparrow}[\tilde\omega_b-(n-1)\chi_{bb}]\mathcal{D}[\hat\sigma_{n-1,n}]\right)\rho \nonumber \\
&+\sum_{n=0}^{\infty}\left(\frac{|g_1|(\chi_{bb}-\Delta)}{(n\chi_{bb}+\Delta)((n-1)\chi_{bb}+\Delta)}\right)^2 \left(\Gamma_{\downarrow}[\omega_{+\Delta,0}]\mathcal{D}[\hat{a}^2\hat\sigma_{n,n}]+\Gamma_{\uparrow}[\omega_{+\Delta,0}]\mathcal{D}[\hat{a}^{\dagger 2}\hat\sigma_{n,n}]\right)\rho\nonumber\\
&+\sum_{n=0}^{\infty}\left(\frac{|g_2|(2\chi_{bb}+\Delta)}{((n-2)\chi_{bb}-\Delta)((n-1)\chi_{bb}-\Delta)}\right)^2 \left(\Gamma_{\downarrow}[\omega_{-\Delta,1}]\mathcal{D}[\hat{a}^2\hat\sigma_{n,n}]+\Gamma_{\uparrow}[\omega_{-\Delta,1}]\mathcal{D}[\hat{a}^{\dagger 2}\hat\sigma_{n,n}]\right)\rho\nonumber\\
&+\sum_{n=0}^{\infty}\left(\frac{\sqrt{(n+1)(n+2)}|g_1|\chi_{bb}}{(n\chi_{bb}+\Delta)((n+1)\chi_{bb}+\Delta)}\right)^2 \left(\Gamma_{\downarrow}[\omega_{+\Delta,2n}]\mathcal{D}[\hat{a}^{\dagger 2}\hat\sigma_{n+2,n}]+\Gamma_{\uparrow}[\omega_{+\Delta,2n}]\mathcal{D}[\hat{a}^2\hat\sigma_{n,n+2}]\right)\rho\nonumber\\
&+\sum_{n=0}^{\infty}\left(\frac{\sqrt{(n+1)(n+2)}|g_2|\chi_{bb}}{(n\chi_{bb}-\Delta)((n-1)\chi_{bb}-\Delta)}\right)^2 \left(\Gamma_{\downarrow}[\omega_{-\Delta,2n+1}]\mathcal{D}[\hat{a}^{\dagger 2}\hat\sigma_{n+2,n}]+\Gamma_{\uparrow}[\omega_{-\Delta,2n+1}]\mathcal{D}[\hat{a}^2\hat\sigma_{n,n+2}]\right)\rho \label{eq:after_RWA_ME}
\end{align}
\end{widetext}
where 
\begin{align*}
\Gamma_{\downarrow}[\omega]&= \frac{4(1+n_{\mathrm{th}}[\omega])|\Omega[\omega]|^2}{\gamma[\omega]}\\
\Gamma_{\uparrow}[\omega]&=  \frac{4 n_{\mathrm{th}}[\omega]|\Omega[\omega]|^2}{\gamma[\omega]}.
\end{align*}
In the next section we study the adiabatic elimination of the junction mode.

\section{Validity of adiabatic elimination}
\label{appendix:optimizing_gamma}

Here we limit ourselves to the lowest three levels ($\ket{g}$, $\ket{e}$ and $\ket{f}$) of the junction mode and furthermore we assume the bath to be at zero temperature. Additionally, following the discussion in Section~\ref{sec:two_photon_dissipation_error}, we consider the case of an engineered bath potentially leading to a strong direct dissipation from $f$ to $g$. The master equation is given by 
\begin{align}
\frac{\mathrm{d}}{\mathrm{d}t}\rho &=-\frac{i}{\hbar}\left[H_{\mathrm{eff}},\rho\right] +\left(\Gamma_\downarrow[\tilde{\omega}_b]\mathcal{D}[\hat\sigma_{eg}]+ 2\Gamma_\downarrow[\tilde{\omega}_b-\chi_{bb}]\mathcal{D}[\hat\sigma_{fe}]\right)\rho \nonumber\\
&+\left(\kappa_{2,gg}\mathcal{D}[\hat{a}^2\hat\sigma_{gg}]+\kappa_{2,ee}\mathcal{D}[\hat{a}^2\hat\sigma_{ee}]+\kappa_{2,ff}\mathcal{D}[\hat{a}^2\hat\sigma_{ff}]\right)\rho\nonumber\\
&+\Gamma_{fg}^\mathrm{eng}\mathcal{D}[\hat\sigma_{fg}]\rho+\kappa_{2,fg}\mathcal{D}[\hat{a}^{\dagger 2}\hat\sigma_{fg}]\rho \label{eq:before_elimination}
\end{align}
where the decay rates $\kappa_{2,gg}$, $\kappa_{2,ee}$, $\kappa_{2,ff}$ and $\kappa_{2,fg}$ can be inferred from \eqref{eq:after_RWA_ME}. The rate $\Gamma_{fg}^{\mathrm{eng}}$ corresponds to the engineered direct dissipation from $f$ to $g$. Assuming $2\Gamma_{\downarrow}[\tilde{\omega}_b-\chi_{bb}]+\Gamma_{fg}^{\mathrm{eng}}\gg \|H_{\mathrm{eff}}/\hbar\|$, we adiabatically eliminate the junction mode to obtain the master equation in \eqref{eq:effective_eom}. Note, however, that for this general noise spectrum, the effective dissipation rates are given by 
\begin{align}
\kappa_{4\mathrm{ph}} =& \frac{4 |g_{4\mathrm{ph}}|^2}{2\Gamma_{\downarrow}[\tilde\omega_b-\chi_{bb}]+\Gamma_{fg}^{\mathrm{eng}}},\nonumber\\
\kappa_{2\mathrm{ph}} =& \frac{|g_1|^2}{\Delta^2}\Gamma_{\downarrow}[\tilde\omega_b+\Delta]+\frac{|g_2|^2}{(\Delta+\chi_{bb})^2}\Gamma_{\downarrow}[\tilde\omega_b-\Delta-\chi_{bb}].\label{eq:diss_gen_spec}
\end{align}
The rates given in~\eqref{eq:kappa_4ph_kappa_2ph} and~\eqref{eq:improved_scheme_rates} correspond to the white noise case where $\Gamma_{\downarrow}[\omega_k]=\Gamma_1$ for all $k$. 

The above general result provides another possible approach to mitigate the problem of the undesired two-photon dissipation. The two dissipation rates $\kappa_{4\mathrm{ph}}$ and $\kappa_{2\mathrm{ph}}$ from \eqref{eq:diss_gen_spec} are sensitive to noise at different frequencies. While $\kappa_{4\mathrm{ph}}$  involves the noise at frequency $\tilde\omega_b-\chi_{bb}$, the undesired $\kappa_{2\mathrm{ph}}$ involves the noise at frequencies $\omega_1 = \tilde\omega_b+\Delta$ and $\omega_2 = \tilde\omega_b-\Delta-\chi_{bb}$. It is possible to engineer the coupling of the system to an electromagnetic bath such that  $\Gamma_{\downarrow}[\tilde\omega_b], \Gamma_{\downarrow}[\tilde\omega_b-\chi_{bb}]\gg \Gamma_{\downarrow}[\omega_1],\gamma_{\downarrow}[\omega_2]$. Indeed, one can mediate the coupling between the system and the bath through a band-pass filter. This is a more elaborate version of the Purcell filter realized in~\citep{Reed2010}. The frequencies $\tilde\omega_b$ and $\tilde\omega_b-\chi_{bb}$ have to be in the pass band, whereas the frequencies $\omega_1$ and $\omega_2$ have to be in the cut-off. 

The rest of this appendix is devoted to checking the validity of this adiabatic elimination through numerical simulations.  In Fig.\ref{fig:adiabatic_validity}a, we compare the dynamics given by \eqref{eq:before_elimination} and \eqref{eq:effective_eom}, using the same parameters as in Fig.\ref{fig:simulation_results} (corresponding to $|\alpha|^2=4$ and $g_{4\mathrm{ph}}/2\pi=\SI{41}{\kilo\hertz}$), and taking $\Gamma_{fg}^{\mathrm{eng}}/2\pi=\SI{0.6}{\mega\hertz}$ (black) and $\Gamma_{fg}^{\mathrm{eng}}/2\pi=\SI{6}{\mega\hertz}$ (magenta). For the considered amplitude $|\alpha|^2=4$, the choice of $\Gamma_{fg}^{\mathrm{eng}}/2\pi=\SI{6}{\mega\hertz}$ satisfies the above separation of time-scales, leading to a good agreement between the dashed and solid magenta lines. The choice of $\Gamma_{fg}^{\mathrm{eng}}/2\pi=\SI{0.6}{\mega\hertz}$ leads to a disagreement with the reduced dynamics. Note that the dynamics still converges towards the expected state albeit at a slower rate. To choose the optimum working point, we perform simulations of~\eqref{eq:before_elimination}, sweeping $\Gamma_{fg}^{\mathrm{eng}}/2\pi$ from $\SI{0.1}{\mega\hertz}$ to $\SI{10}{\mega\hertz}$. In Fig.\ref{fig:adiabatic_validity}b we plot the overlap with the $\cZero$ cat state as a function of time and $\Gamma_{fg}^{\mathrm{eng}}$. From this, we extract the time taken to achieve 90\% fidelity as a function of $\Gamma_{fg}^{\mathrm{eng}}$.  This corresponds to the green curve in~Fig.\ref{fig:adiabatic_validity}c. As illustrated by the green dot, the optimum working point is given by $\Gamma_{fg}^{\mathrm{eng}}/2\pi=\SI{2}{\mega\hertz} $. Note that the working point of $\Gamma_{fg}=\SI{4}{\mega\hertz}$ used in Fig.~\ref{fig:simulation_results} is selected to be well in the region of adiabatic validity while still getting a strong four-photon dissipation ($\kappa_{4\mathrm{ph}}$). The same simulations for $|\alpha|^2=3$ and $5$ give rise to different working points at $\SI{1}{\mega\hertz}$ and $\SI{3.7}{\mega\hertz}$ respectively. Indeed, the norm $\|H_{\mathrm{eff}}\|$, in the assumption $2\Gamma_{\downarrow}[\tilde{\omega}_b-\chi_{bb}]+\Gamma_{fg}^{\mathrm{eng}}\gg \|H_{\mathrm{eff}}/\hbar\|$, corresponds to the norm of the Hamiltonian when confined to the code space $\textrm{span}\{\ket{\pm\alpha},\ket{\pm i \alpha}\}$. This implies a separation of time-scales which depends on the amplitude $|\alpha|$ of the cat state, therefore leading to different optimum working points.



\end{document}